\documentclass[a4paper, 11pt]{llncs}
\usepackage{amsmath,amssymb}
\usepackage{graphicx}
\usepackage{subfig}
\usepackage{cite}
\usepackage{url}

\urldef{\email}\path|{aysajan, jalar}@isy.liu.se|
\newcommand{\keywords}[1]{\par\addvspace\baselineskip
\noindent\keywordname\enspace\ignorespaces#1}

\newcommand{\mZ}{\mathcal{Z}}

\newcommand{\mH}{\mathcal{H}}
\newcommand{\mM}{\mathcal{M}}
\newcommand{\mT}{\mathcal{T}}

\renewcommand{\epsilon}{\varepsilon}

\begin{document}

\mainmatter  

\title{Security of Authentication with a Fixed Key in Quantum Key Distribution}
\titlerunning{Security of authentication with a fixed key \ldots}

%
%
\author{Aysajan Abidin
\and Jan-{\AA}ke Larsson}
\authorrunning{A. Abidin and J.-{\AA}. Larsson}

\institute{Department of Electrical Engineering, \\
           Link\"oping University, SE-581 83 Link\"oping, Sweden\\
\email}
%
%

\maketitle
\begin{abstract}
  We study the security of a specific authentication procedure of
  interest in the context of Quantum Key Distribution (QKD). It works
  as follows: use a secret but fixed Strongly Universal$_2$ (SU$_2$)
  hash function and encrypt the output tag with a one-time pad
  (OTP). If the OTP is completely secret, the expected time for an
  adversary to create a tag for a chosen message is exponential in the
  tag length. If, however, the OTP is partially known in each
  authentication round, as is the case in practical QKD protocols,
  then the picture is different; the adversary's partial knowledge of
  the OTP in each authentication round gives partial information on
  the secret hash function, and this weakens the authentication in
  later rounds.  The effect of this is that the lifetime of the system
  is linear in the length of the fixed key.  This is supported by the
  composability theorem for QKD, that in this setting provides an
  upper bound to the security loss on the secret hash function, which
  is exponential in the number of authentication rounds.  This needs
  to be taken into account when using the protocol, since the authentication 
  gets weakened at each subsequent round and thus the QKD generated is key 
  is not as strong as when the authentication is strong. Some
  countermeasures are discussed at the end of this paper.
\end{abstract}

\section{Introduction}

QKD is a provably secure key growing
technique based on the laws of quantum physics. It was first
introduced by Bennett and Brassard in 1984 \cite{BB84}, and uses a
so-called quantum channel that obeys the laws of quantum physics,
together with a public communication channel.  A QKD round consists of
five steps: raw key generation on the quantum channel, followed by
sifting, error detection and reconciliation, privacy amplification,
and authentication on the public channel; see \cite{BBBSS, BS, BBR,
  BBR2, BBCM, Lutken} for the details of these steps.  Practical
implementations of QKD need a low-noise quantum channel but also an
immutable public communication channel. Without the latter, QKD can trivially 
be broken by a man-in-the-middle attack. Therefore, secure message
authentication is indispensable for the security of QKD \cite{SML}.

In the standard proposed QKD, authentication is achieved by using the
Wegman-Carter approach \cite{WC,WC2}, based on the idea of Universal
hashing. The security of the Wegman-Carter authentication in the
context of QKD was studied in \cite{CL}, noting some problems arising
from usage of a partially known key, and detailing some
countermeasures.

\subsection{Authentication with secret fixed hash function and OTP}

The main goal of this paper is to study the security of authentication
with a fixed key in the context of QKD. Namely, we study the security
of an authentication procedure that works as follows: The legitimate
communicating parties, Alice and Bob, share a secret but fixed hash
function $f$ taken at random from a SU$_2$ hash function family and a
short secret key to be used as OTP in advance. During the public
discussion phase of each QKD round, Alice sends the classical message
and tag pair $m+t$ with $t = f(m) \oplus K$, where $K$ is an OTP, to
Bob. Upon receiving the message-tag pair $(m,t)$, Bob verifies whether
the message $m$ did originate from Alice by comparing $f(m) \oplus K$
to $t$: If they are identical, then he accepts $m$ as authentic;
otherwise, he rejects it.

This authentication primitive was originally proposed by Wegman and
Carter in \cite{WC2} with the intent to reduce the key consumption
rate of authentication.  Low key consumption is essential in QKD,
since the key consumption rate of the used authentication directly
influences the key growing rate. Wegman-Carter authentication using an
$\epsilon$-ASU$_2$ hash function family has a key consumption rate
which is logarithmic in the message length, while using encrypted tags
would reduce this; the rate is linear in the tag length as the round
count increases.

Partial knowledge of the OTP key $K$ leaks information on secret but
fixed SU$_2$ hash function $f$. In QKD, the privacy amplification step
reduces the information leaked to Eve during each round, but not all
the way to zero.  Thus Eve may still have some partial knowledge of
the OTP key used for authentication in the subsequent rounds. This
information, $\epsilon$, on the OTP key $K$ in each round leaks
$\epsilon$ information on the secret hash function $f$.  Intuitively,
the information on $f$ leaked to Eve is linear in the number of
authentication rounds.  In what follows, we show that this is really
the case, and in fact Eve will eventually have enough knowledge of the
hash function $f$ to enable her to create a tag for her forged
message. Furthermore, the composability theorem for QKD gives an
exponential upper bound for the security loss of the system.

\subsection{Our contributions}

In the case when the OTP key $K$ is completely secret to Eve, it
behaves as an evenly distributed random variable to her (which is the
reason for the upper-case $K$ notation). In this case, the best attack
for Eve would be to guess the value of $t_\text E$, the tag value for
her message $m_\text E$.  Since all tag values are equally possible,
the probability of each guess succeeding is one divided by the size of
all possible tags $|\mT|$. Furthermore, she can gain no knowledge
about the secret hash function $f$ from guessing, because $K$ in the
current round is independently distributed from previous rounds.  The
probability of a successful guess would in each round be
$1/|\mT|=2^{-\log |\mT|}$, which implies that the expected lifetime
\begin{equation}\label{eq:guess}
n = |\mT| = 2^{\log |\mT|} 
\end{equation}
is exponential in the tag length $\log|\mT|$. 

We are interested in seeing how this exponential lifetime behavior
would change if Eve has some knowledge of $K$ in each round. In the
remainder of this paper we estimate how many rounds it takes for Eve
to gain complete knowledge of the secret but fixed hash function $f$
(taken at random from an SU$_2$ family), under the assumption that the
practical implementation of QKD protocol generates $\epsilon$-perfect
key in each run. We refer to \cite{Renner} for the definitions of
perfect and $\epsilon$-perfect keys, and of ideal and $\epsilon$-ideal
protocols. Note that since the authentication primitive uses a fixed
SU$_2$ hash function, the sequence of the security parameters for the
key stream generated from the QKD protocols \emph{cannot} be made a
geometric sequence by increasing the protocol complexity at each run,
as discussed in \cite{Renner}. By fixing Eve's partial knowledge of
the OTP key in each authentication round, we derive an estimate for
the lifetime of system which is linear in the length of the key
identifying $f$ and proportional to her partial knowledge of OTP.

This is not in conflict with the composable security of QKD, which
implies that the key generated by QKD can be used securely in
classical cryptographic tasks such as authentication
\cite{Renner,BenOr}.  In this case, however, the authentication
procedure itself degrades as the authentication round count increases.
Below, we show that the composability theorem for QKD predicts that
the security loss on the fixed secret hash function is exponentially
upper bounded in the number of authenticaton rounds.

It should be pointed out that the attack needs a large computational
capacity of the attacker. Usually, no bounds are imposed on the
computational capacity of an eavesdropper attacking a QKD system. This
is because QKD is provably secure based on laws of nature, rather than
computational complexity as is usually the case for key-sharing
systems.  This large computational need of the attack will
unfortunately limit simulations in this paper because of our bounded
computational power.

\subsection{Organization of the paper}

The rest of the paper is organized as follows. In Section 2, we
present an attack and estimate its effect on the system under
simplifying assumptions, and also present simulations on a SU2 hash
function family, followed by a modification to the attack that
establishes the desired lifetime. Section 3 contains the theoretical
upper bound for the security loss predicted by the composability
theorem for QKD.  Finally, Section 4 concludes the paper.

\textbf{Notation.} In what follows, $\mM$ is the set of messages,
$\mT$ is the set of tags, $\mH$ is a family of hash functions
$f:\mM\to\mT$ with $|\mH|=H+1$, and $\mH_i$ are integer-indexed
subsets of $\mH$.  Logarithms are in base 2. The random variables used
are $K$, $N$, and $X_i$, while lower-case $m$ denotes a message and
$t$ a tag, $m_\text E$ and $t_\text E$ are Eve's message and attempt
at a tag.

\section{Attack and lifetime estimate}

Eve would like to perform an attack which is better than simply
guessing the tag. Ideally, it should be better in two ways: it should
succeed with high probability, and should not be detected easily. Eve
wants in essence a good covert attack. The below description
delineates an attack which achieves both goals: the expected number of
rounds until success will be much lower than for the guessing attack,
and in addition, the attack is covert, meaning that Eve only listens
to the communication between Alice and Bob for a number of rounds, and
only launches an attack when she is sure that it will succeed.

The attack is as follows: Eve's goal is to identify the used hash
function $f$ among the $H+1$ hash functions in $\mH$, i.e., to
eliminate $H$ functions from $\mH$. In each QKD round, Eve intercepts
a valid (classical) message-tag pair $m+t$, where $t=f(m)\oplus K$,
from, say, Alice to Bob.  The random variable $K$ (random to Eve) is
not entirely evenly distributed because of Eve's partial knowledge. We
will, in what follows, assume that her knowledge is such that she
knows a few values of $K$ that has probability 0.  She uses this
knowledge to identify possible candidates for $f(m)$.  This means that
in each run, Eve can identify a subset $\mH_i$ out of all the possible
hash functions in $\mathcal H$ by eliminating the hash functions (in
$\mH$) that do not hash $m$ to the set of possible candidates for
$f(m)$. The set $\mH_i$ will consist of the true match (the fixed
secret hash function) and a number of false matches.

The set $\mH_i$ can in principle be of different size depending on the 
hash function family, which hash function is used, and the message, but 
here we are focusing on Strongly Universal hash function families and in 
this case, the inverse image of any tag has the same size, and each 
subset has the same size $|\mH_i| = h$. Therefore, Eve's information on $K$ 
in terms of min-entropy translates directly into the quantity $-\log(h/H)$. 

\subsection{Bounds using simplifying assumptions}

After $i$ runs the set of possible hash functions will decrease to
$\cap_{j=1}^i\mathcal H_j$.  In general, the remaining number of false
matches in this intersection is a random variable
\begin{equation}
X_i=|\cap_{j=1}^i\mathcal{H}_j|-1. 
\end{equation}
We are interested in the expected time it takes until Eve has
identified the (no longer secret) true hash function, that is, the
expectation of the (random) index $N$ that is the earliest that gives
$X_N = 0$ (such that $X_{N-1}\ge 1$). 

By assuming that that each round is independent of the former, and
that each subset is exactly evenly distributed within the previous
subset, we obtain $X_i=X_{i-1}h/H$. This is only possible when the
$X_i$ are continuous variables; we will analyze the discrete
(integer-valued) case below.  With $X_0=k$ we obtain $X_1=kh/H$,
$X_2=k(h/H)^2$, \ldots, $X_l=k(h/H)^l$. Now, our demand
$(X_N=0)\cap(X_{N-1}\ge1)$ translates into $(X_N<1)\cap(X_{N-1}\ge1)$,
which in turn implies that $N|(X_0=k)$ is not random in this case, but
is in fact equal to a number $n_k$ for which
\begin{equation}
  \label{eq:2}
  k(\tfrac hH)^{n_k}<1\le k(\tfrac hH)^{n_k-1},
\end{equation}
which after some algebra simplifies to
\begin{equation}
  \label{eq:3}
  n_k-1\le\frac{\log k}{-\log \tfrac hH}< n_k,
\end{equation}
that is,
\begin{equation}
  \label{eq:4}
  n_k=\left\lceil \frac{\log k}{-\log \tfrac hH}\right\rceil.
\end{equation}
In particular, $n_H=\lceil \log H/(-\log (h/H))\rceil $, which means
that the lifetime of the system would be directly proportional to the
key length\footnote{Here, the length of the key identifying the secret
  hash function is actually $\log(H+1)$.} divided by the information
on the OTP used in each step.  This is what we would expect of a
system in which there is a constant gain of information in each run.

In the discrete case, the analysis is more complicated. We extend to a
random draw of hash functions, but keep the assumption that each round
is independent of the former. This means that the probability of
drawing a hash function present in $\cap_{j=1}^{i-1} \mH_j$ in run $i$
only depends on $X_{i-1}$, which corresponds to a random draw of $h$
elements without replacement from $\mathcal H$, where there are two
types of elements: those in $\cap_{j=1}^{i-1}\mathcal H_j$ ($X_{i-1}$
of them), and those outside the set.  In other words, the number of
hash functions in $\cap_{j=1}^i\mathcal H_j$ given $X_{i-1}$ is
hypergeometrically distributed, so that,
\begin{equation}
  \label{eq:6}
  p_{jk}:=P(X_i=j|X_{i-1}=k)=\frac{{k \choose j}{H-k \choose h-j}}
  {{H  \choose h}}.
\end{equation}
The expected lifetime time when $k$ false hash functions remain is
\begin{equation}
  \label{eq:12}
  n_k=E(N|X_0=k).
\end{equation}
Then, $n_0=0$ and
\begin{equation}
  \label{eq:14}
  \begin{split}
    n_k&=\sum_{j=0}^k E(N|X_1=j)P(X_1=j|X_0=k)\\
    &=\sum_{j=0}^k \Big(E(N|X_0=j)+1\Big)P(X_1=j|X_0=k)=1+\sum_{j=0}^k p_{jk}n_j.
  \end{split}
\end{equation}
Solving for $n_k$ gives
\begin{equation}
  \label{eq:15}
  n_k=\frac{1+\sum_{j=0}^{k-1} p_{jk}n_j}{1-p_{kk}},
\end{equation}
and since $p_{jk}, \, j=0,1,\cdots,k$, are given explicitly above, the
$n_k$ can be calculated from this equation, although the expressions
are complicated.  

The goal is to prove a logarithmic bound on $n_k$ in terms of $k$ as in \eqref{eq:4} in general. 
Splitting the sum of \eqref{eq:14} at the point $s$ (just before the index $\lceil s\rceil$) 
gives 
\begin{equation}
 \begin{split}
  \label{eq:27}
    n_k&=1+n_{\lceil s\rceil -1}P(X_i<s | X_{i-1}=k) + n_k P(X_i>s | X_{i-1}=k).
  \end{split}
\end{equation}
And now solving for $n_k$ gives
\begin{equation}
  \label{eq:28}
  n_k\le\frac1{1-P(X_i>s |X_{i-1}=k)}+n_{\lceil s\rceil -1}. 
\end{equation}
If the probability in the denominator does not grow to fast when $s$ decreases 
from $k-1$, we can use a value $s$ sufficiently far from $k$ to establish logarithmic 
growth of $n_k$ in $k$. The one-sided Chebyshev inequality implies 
\begin{equation}
  \label{eq:29}
  \begin{split}
    P\Big((X_i>s\Big|X_{i-1}=k\Big)\le
    \frac{V\big(X_i\big|X_{i-1}=k\big)}{(s-E(X_i|X_{i-1}=k))^2+V\big(X_i\big|X_{i-1}=k\big)},
  \end{split}
\end{equation}
so that 
\begin{equation}
  \label{eq:31}
  \begin{split}
    &\frac1{1-P\Big(X_i\ge s\Big|X_{i-1}=k\Big)}
    \le \frac {\Big(s-E\big(X_i\big|X_{i-1}=k\big)\Big)^2 + V\big(X_i\big|X_{i-1}=k\big)}
    {\Big(s-E\big(X_i\big|X_{i-1}=k\big)\Big)^2}\\ 
    &\quad = 1+\frac {V\big(X_i\big|X_{i-1}=k\big)}
    {\Big(s-E\big(X_i\big|X_{i-1}=k\big)\Big)^2}
    \le 1 +\frac {k\frac hH\Big(1-\frac hH\Big)}
    {\Big(s-k\frac hH\Big)^2}.
  \end{split}
\end{equation}
This implies that 
\begin{equation}
  \label{eq:33}
  \begin{split}
    n_k&\le1 + \frac {k\frac hH\Big(1-\frac hH\Big)}
    {\Big(s-k\frac hH\Big)^2} + n_{\lceil s\rceil -1}.
  \end{split}
\end{equation}
Note that even if $\lceil s\rceil$ and $k$ coincide, the 
indices above do not. Now let us prove using induction that
\begin{equation}
  \label{eq:34}
  n_k\le a+b\log k
\end{equation}
with the appropriate $a$ and $b$. A simple starting point is $n_1$:
\begin{equation}
  \label{eq:35}
  a=n_1=\frac1{1-\frac hH}.
\end{equation}
Now, we assume \eqref{eq:34} holds for $k$ less than $p\ge 2$ which implies  
\begin{equation}
  \label{eq:36}
  n_{\lceil s\rceil -1}\le a + b\log(\lceil s\rceil - 1) \le a + b\log s = b\log\frac{s}{k}+ a 0+ b\log k,
\end{equation}  
so that
\begin{equation}
  \label{eq:37}
  n_p \le 1 + \frac {k\frac hH\Big(1-\frac hH\Big)}{\Big(s-k\frac hH\Big)^2} 
            + b\log\frac{s}{k} + a + b\log k.
\end{equation}
Choosing
\begin{equation}
  \label{eq:38}
  b=-\left(1+\frac {k\frac hH\Big(1-\frac hH\Big)}{\Big(s-k\frac hH\Big)^2}\right ) 
  \Big / \log\frac{s}{k}>0
\end{equation}
gives the desired 
\begin{equation}      
  n_p\le a+b\log p.
  \label{eq:39}
\end{equation}
By induction we obtain that the lifetime $n_k$ is bounded by 
\begin{equation}
  \label{eq:40}
  n_k\le \frac{1}{1-\frac{h}{H}}+\left(1+\frac {k\frac hH\Big(1-\frac hH\Big)}
    {\Big(s-k\frac hH\Big)^2}\right)\frac{\log k}{-\log\frac{s}{k}}.
\end{equation}

If $s$ is chosen proportional to $k$, the first term in the parenthesis will 
dominate at large values of $k$, and the proportionality constant appears in 
$\log s/k$. Choosing $s=k\sqrt{h/H}$, then with similar calculations as above we obtain
\begin{equation}\label{eq:41}
n_k \le \frac{1}{1-\frac{h}{H}} + \left(1+\frac{1+\sqrt{\frac{h}{H}}}{k(1-\sqrt{\frac{h}{H}})}\right)
\frac{2\log k}{-\log\frac{h}{H}},
\end{equation}
where the coefficient in front of the logarithm decreases to 1 when
$k$ increases.  The bound for $n_H$ is logarithmic in $H$ and slightly
larger than the one in \eqref{eq:4}, which is natural taking the
broadening of the distribution into account. 
This is similar as in the previous more simplified situation; the
lifetime of the system is linear in the key length rather than
exponential in the tag length. We now need to check the remaining
assumption that each round is independent of the former: does the
random draw in each round follow a hypergeometric distribution?

\subsection{Simulations for an SU$_2$ Family}
 
We want to simulate an authentication system with a secret fixed hash
function from a SU$_2$ hash function family, where the tag is OTP
encrypted with a partially known key. Here, we restrict ourselves to a
specific hash function family from \cite{WC} as follows.  Let $\mM$
and $\mT$ be finite sets of messages and tags, respectively.  Let $p$
be smallest prime number greater than $|\mM|$. For each integer
$0<q<p$ and $0\le r <p$, define a hash function $f_{(q,r)}:\mM\to\mT$
by
\begin{equation}
\label{eq:H1}
f_{(q,r)}(m) \equiv ((mq + r) \mod{p}) \mod{|\mT|}.
\end{equation}
Then, $\mH_1 = \{f_{(q,r)}: q\in\mZ_p\setminus\{0\} 
\mbox{ and } r\in\mZ_p\}$ is an SU$_2$ hash function family. 
This family was introduced as "$\mH_1$" in \cite{WC} (the index 
is not used in the same way as in this paper), and is not quite 
SU$_2$, in Wegman and Carter's own words: it is "close".  

The parameters chosen for our simulations will admittedly be very
restrictive and somewhat unrealistic when compared to a full-blown QKD
system.  The reason for this is our bounded computational
capabilities; as already mentioned, no bound is usually imposed on an
attacker in QKD, but this does unfortunately not apply to authors of
scientific papers. The largest hash function family we will use will
be of size $2^{28}$, and our attack uses the equivalent of
round-by-round exhaustive search, by keeping track of eliminated keys
at each round, and this gives a high computational demand.  The hash
function family size will not be kept fixed in the different
simulations. We use a set $\mT$ of tags with size $2^7$, and message
sets $\mM$ with a varying size from $2^9$ through $2^{13}$. For each
pair of $\mM$ and $\mT$, there is a corresponding hash function family
$\mH$. We set Eve's partial information on the OTP key $K$ to $10\%$,
again an unrealistically high number, but this is chosen to show the
results qualitatively while still bounding the lifetime of the system,
see below.

\begin{figure}[ht!]
  \center
  \includegraphics[width=8cm]{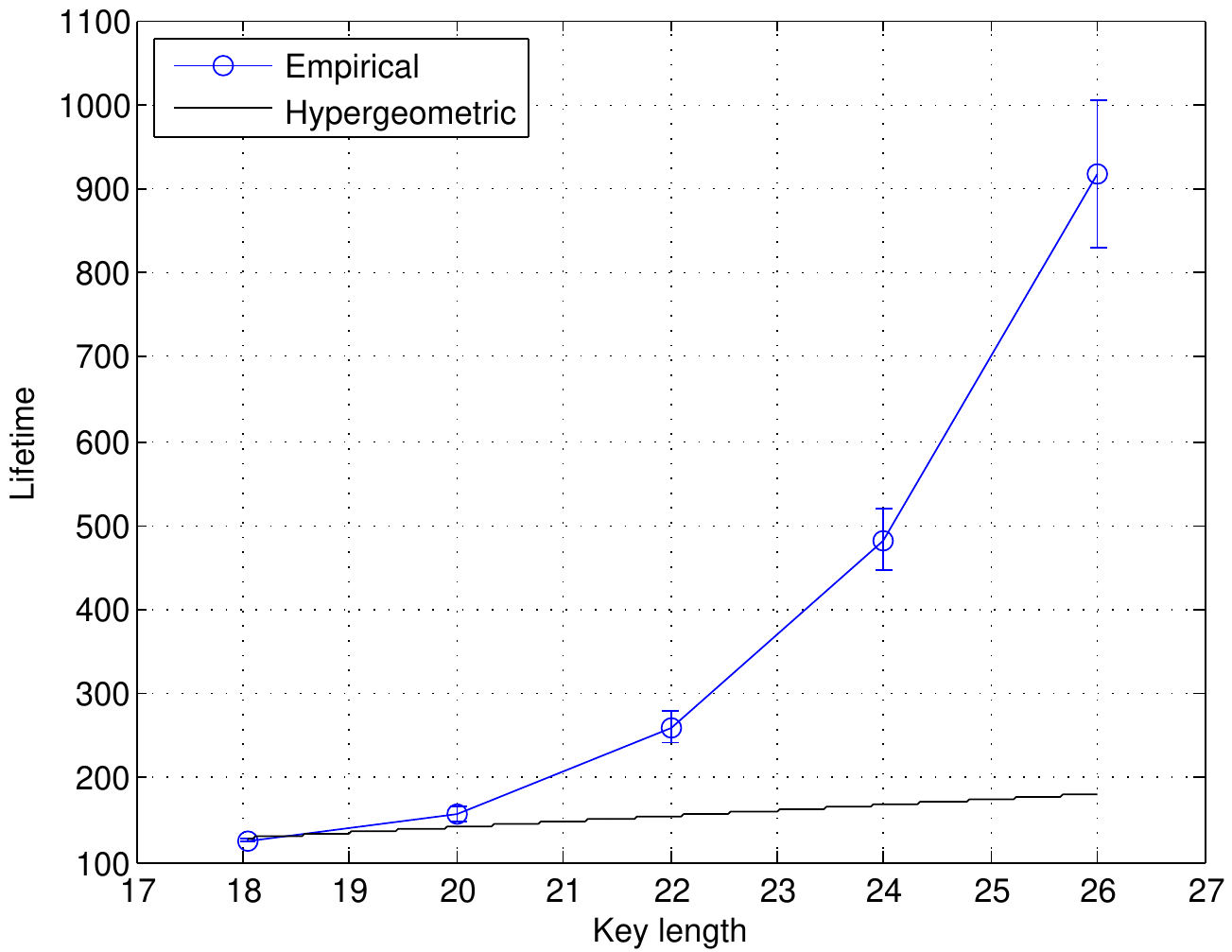}
  \caption{The number of rounds until the secret hash function $f$ is
    found when it is taken at random from the family $\mH_1$.}
  \label{fig:LT}
\end{figure}

The simulations are done as follows: a hash function $f$ is taken at
random from the appropriate SU$_2$ family. In each round, a message
$m_i$ is randomly drawn, and the tag $t_i$ is calculated using
$f$. This tag is entered into the set $\mT_i$, and more tags are
randomly chosen from $\mT$ to make $|\mT_i|=h|\mT|/|\mH|$, which
corresponds to a situation where Eve can use the OTP-encrypted tag
$t_i\oplus K$ together with her partial knowledge of $K$ to identify
the set $\mT_i$. She then uses this set to identify the set of
possible hash functions $\mH_i$, and she forms the intersection
$\cap_{j=1}^i\mathcal H_j$. When the intersection has been reduced to
just one hash function, Eve has identified $f$, and this is repeated
many times to estimate the lifetime of the system, the results can bee
seen in Fig.~\ref{fig:LT}.

As we can see, the lifetime is not as was estimated in \eqref{eq:41}.
It now increases exponentially as the key length increases, contrary
to our earlier linear estimate. The reason for this is that the rounds
are not independent, at least not for this hash function family. This
is especially pronounced when there are few hash functions left: most
of the increase occurs when waiting for the last few false matches to
disappear. Recall that the hash functions are eliminated by using the
inverse image, for one message in each round from a set of
``possible'' tags, to a set of ``possible'' hash functions. And hash
functions that have not been eliminated already have a lower
probability to be eliminated than they would in the case of
independent rounds.

However, Eve's goal is not really finding the secret hash function
$f$. Eve's objective is to be able to generate the correct tag for her
(forged) message, to breach security of the authentication.  So far,
our focus has been on finding the secret hash function $f$.  We note
that even if the remaining set $\cap_{j=1}^i\mathcal H_j$ contains
more than one hash function, Eve can generate the correct
(unencrypted) tag for her message if all the remaining hash functions
map her message to the same value (say, $t_\text E$). Eve can check
for this event, by comparing tags for her message for the different
remaining hash functions. When there are few hash functions
remaining, and they have a low probability to be eliminated, the
probability is high that a random message is mapped to the same tag by
all remaining hash functions. This means that the probability for
Eve's message to be mapped to the same tag is high.

\begin{figure}[ht!]
\begin{center}
\includegraphics[width=8cm]{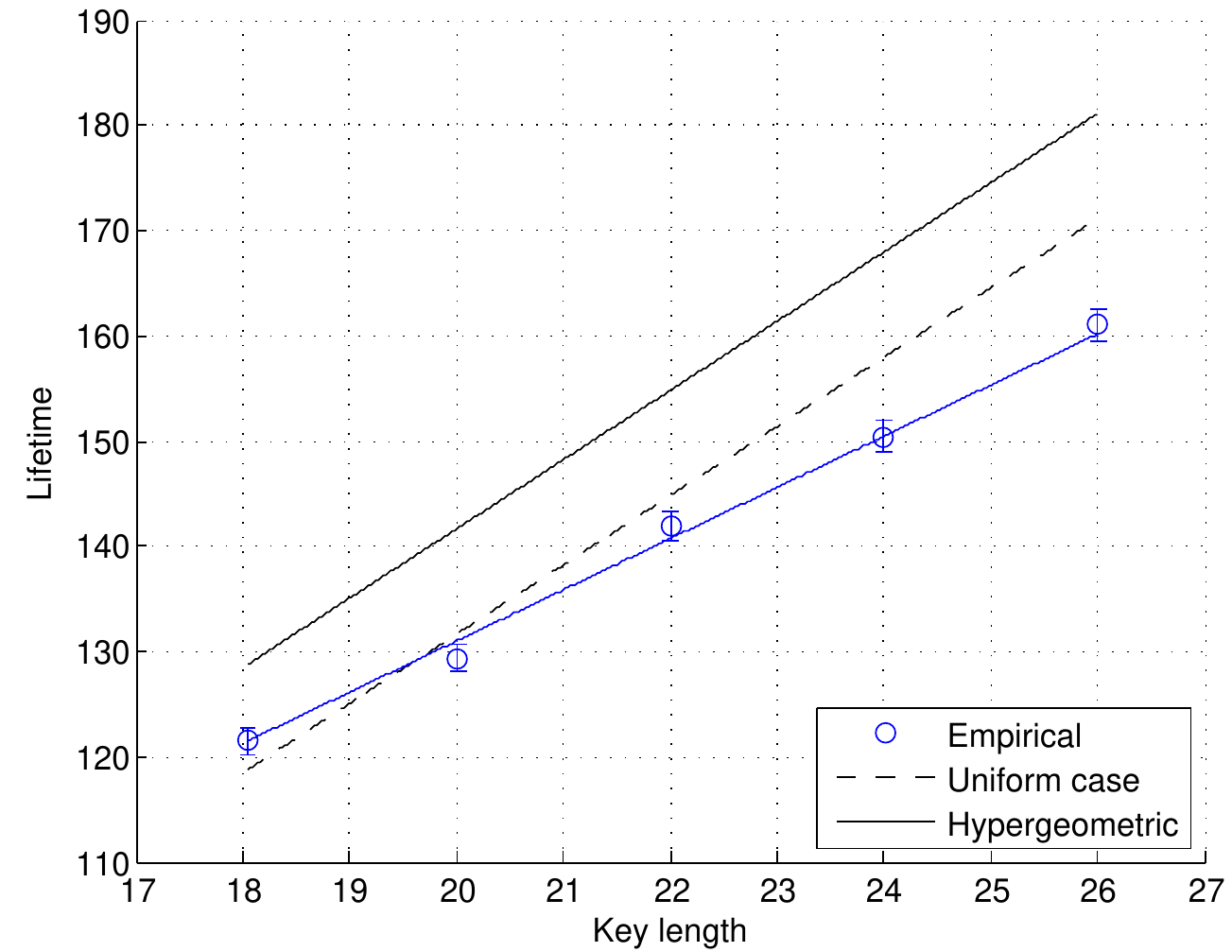}
\caption{The number of rounds until $f$ is found under the uniform and
  hypergeometric assumptions, and the number of rounds until Eve gains
  enough information to generate the valid tag for her forged
  message.}
\label{fig:unihyp}
\end{center}
\end{figure}

Eve also needs to identify the OTP to encrypt her tag. She can do that
when the remaining hash functions in $\cap_{j=1}^i\mathcal H_j$ also
map Alice's message to the same value $t$ (possibly different from
$t_\text E$).  Using the value of $t$, Eve can identify the OTP key
$K$ used, and use that to encrypt her tag.  At this point the system
is broken. Changing the simulation so that Eve checks for this event,
gives a linear lifetime in the key length, as can be seen from
Fig.~\ref{fig:unihyp}.  The simulated lifetime is slightly shorter
than the estimated value, but Eve is solving a simpler task by not
trying to identify the correct hash function $f$ but instead a subset
that has the desired properties.

\section{Upper bound to security loss}

This is not in conflict with composable security of QKD
\cite{Renner,BenOr}. Moreover, the composability can be used to
provide an upper bound to the security loss on the fixed secret hash
function.  The composability theorem for QKD states that if an
$\epsilon_1$-ideal QKD protocol is composed with an $\epsilon_2$-ideal
cryptographic application, e.g., $\epsilon_2$-ideal authentication,
the whole system is $\epsilon_1+\epsilon_2$-ideal. So, if an
$\epsilon_1$-ideal QKD is composed with an $\epsilon_2$-ideal
authentication, then the whole system becomes
$\epsilon_1+\epsilon_2$-ideal and generates an
$\epsilon_1+\epsilon_2$-perfect key. It was argued in \cite{Renner}
that the security parameter for the key stream generated from the
repeated use of QKD can be made arbitrarily small by increasing the
communication complexity of the protocol; that is, by making the
sequence of security parameters for QKD-generated keys a geometric
sequence. This unfortunately is not possible with the authentication
under consideration here. The present authentication procedure uses a
fixed secret SU$_2$ hash function, and this fixes the length of the
message that can be authenticated. Thus, it is reasonable to assume
that the practical implementation of the QKD is $\epsilon_1$-ideal at
a constant $\epsilon_1$ in each round.

\begin{figure}[htp!] 
\begin{center}
\includegraphics[width=8cm]{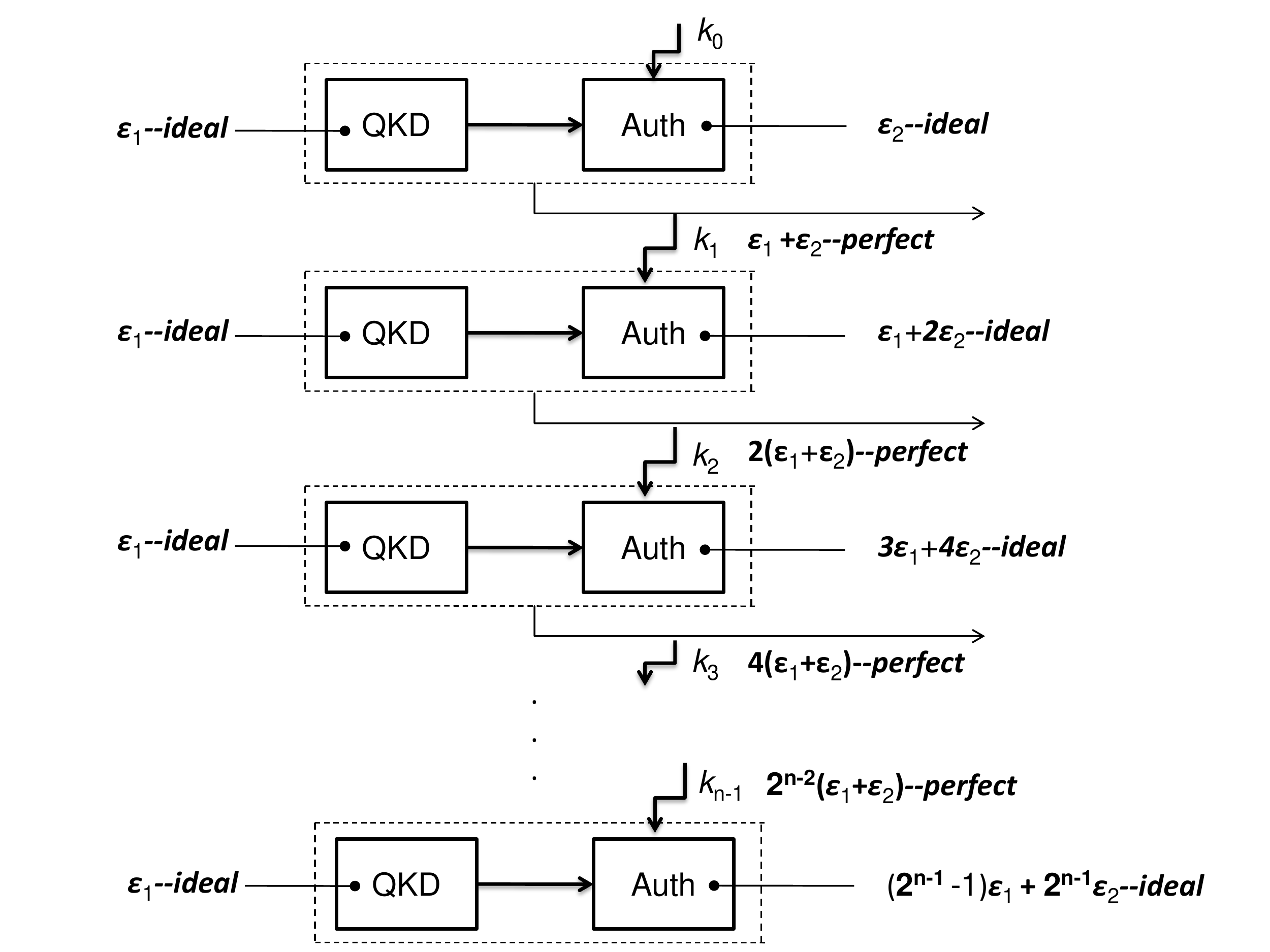}
\end{center}
\caption{Composability diagram of QKD with authentication with fixed key followed by an OTP.}
\label{fig:comp}
\end{figure}

Now, let us look at the sequence of security loss on the secret but
fixed hash function $f$ with help of the composability theorem (see
Fig.~\ref{fig:comp}).
\begin{itemize}
\item In the first round, the composed system of $\epsilon_1$-ideal
  QKD and $\epsilon_2$-ideal authentication produces an
  $\epsilon_1+\epsilon_2$-perfect key. A portion of this
  $\epsilon_1+\epsilon_2$-perfect QKD-generated key will be used as
  the OTP key for the $\epsilon_2$-ideal authentication in the second
  round.
\item In the second round, the composed system of $\epsilon_1$-ideal
  QKD and $\epsilon_2$-ideal authentication using an
  $\epsilon_1+\epsilon_2$-perfect key produces an
  $2(\epsilon_1+\epsilon_2)$-perfect key.  A portion of this
  $2(\epsilon_1+\epsilon_2)$-perfect key will be used as the OTP key
  for the authentication in the third round. Furthermore, the
  $\epsilon_1+\epsilon_2$ information on the OTP key leaks
  $\epsilon_1+\epsilon_2$ information on the fixed hash function,
  which makes the authentication 
  $\epsilon_1+2\epsilon_2$-ideal.
\item In the third round, the composed system of $\epsilon_1$-ideal
  QKD and $\epsilon_1+2\epsilon_2$-ideal authentication using an
  $2(\epsilon_1+\epsilon_2)$-perfect key produces an
  $4(\epsilon_1+\epsilon_2)$-perfect key.  A portion of this
  $4(\epsilon_1+\epsilon_2)$-perfect key will be used as the OTP key
  for the authentication in the fourth round. Furthermore, the
  $2(\epsilon_1+\epsilon_2)$ information on the OTP key leaks
  $2(\epsilon_1+\epsilon_2)$ information on the fixed hash function,
  which makes the authentication 
  $3\epsilon_1+4\epsilon_2$-ideal.
\item In the fourth and following rounds, the process continues,
  doubling the coefficient for each round.
\end{itemize}  
The important property of this authentication scheme is that the
information gained on the fixed hash function $f$ at the current round
carries through to the next round. In other words, the information
leakage on $f$ at each round can be combined.  Therefore, after the $n$-th round, the
information leaked to Eve on the secret but fixed hash function is
$(2^{n-1}-1)\epsilon_1+2^{n-1}\epsilon_2$ so that the authentication
becomes $(2^{n-1}-1)\epsilon_1+2^{n-1}\epsilon_2$-ideal.  The attack
in Section 2 only assumes that the QKD generated key in each round 
is equally strong; in other words, Eve's knowledge of the QKD generated key 
in each round is the same. 

\section{Conclusions}

In this paper, the security of a specific authentication primitive is
studied, a primitive that uses a fixed secret hash function followed
by a one-time-pad encryption on the tag. This is of interest in QKD
because of its low consumption of secret key. We found that, by fixing
Eve's partial knowledge of the OTP key in each QKD round, the lifetime
of the system is linear in the length of the fixed key. Moreover,
using the composability theorem, we found that the leakage of
information on the secret but fixed key is exponentially upper bounded
in the number of authentication rounds.  A suitable countermeasure
would be to change the fixed secret key regularly, at an interval that
ensures that Eve's collected information on the fixed key does not
become too large. This would make the key consumption rate again
logarithmic in the message length, but at a rate much lower than the
standard Wegman-Carter authentication that uses a new
$\epsilon$-ASU$_2$ hash function in each round.

\end{document}